\documentclass[%
 superscriptaddress,
 aip,
 amsmath,amssymb,
 reprint,%
longbibliography
]{revtex4-2}
\usepackage[utf8]{inputenc}
\usepackage[english]{babel}
\usepackage{color} 
\usepackage{graphicx} 
\usepackage{amsmath,amsthm,amsfonts,amssymb,amscd}
\usepackage{calc} 
\usepackage{array} 
\usepackage{titlesec} 
\usepackage{enumitem} 
\usepackage{changepage} 
\usepackage{hyperref} 
\usepackage{miller} 
\usepackage{soul} 
\usepackage{setspace} 
\usepackage{lipsum}
\usepackage{letltxmacro}  
\usepackage{float} 

\newcommand\mytitle{Sampling metrics for robust reconstructions in multislice ptychography: Theory and experiment}

\definecolor{linkColor}{rgb}{0.7,0,0}
\definecolor{darkred}{rgb}{0.7,0,0}

\hypersetup{
    pdftitle={\mytitle},
    pdfauthor={Colin Gilgenbach},
    pdfpagemode=FullScreen,
    pdfborder={0 0 0},
    colorlinks=true,
    urlcolor=linkColor,
    citecolor=linkColor
}

\renewcommand{\AA}[0]{\text{\normalfont\r{A}}}
\newcommand{\code}[1]{\texttt{#1}}
\newcommand{\unit}[1]{\operatorname{#1}}

\titleformat{\section}
{\color{darkred}\sffamily\bfseries}
{\color{darkred}\thesection}{1em}{}
\titleformat{\subsection}
{\color{darkred}\sffamily\itshape}
{\color{darkred}\thesubsection}{1em}{}
\titlespacing*{\section}
{0pt}{8pt}{0pt}
\titlespacing*{\subsection}
{0pt}{8pt}{0pt}

\emergencystretch=\maxdimen
\LetLtxMacro{\ORIGselectlanguage}{\selectlanguage}
\makeatletter
\DeclareRobustCommand{\selectlanguage}[1]{%
  \@ifundefined{alias@\string#1}
    {\ORIGselectlanguage{#1}}
    {\begingroup\edef\x{\endgroup
       \noexpand\ORIGselectlanguage{\@nameuse{alias@#1}}}\x}%
}
\newcommand{\definelanguagealias}[2]{%
  \@namedef{alias@#1}{#2}%
}
\makeatother

\definelanguagealias{en}{english}
\definelanguagealias{EN}{english}
\definelanguagealias{de}{english}

\begin{document}

\title[Sampling Metrics in Multislice Ptychography]{\mytitle}

\author{Colin Gilgenbach}
\affiliation{Massachusetts Institute of Technology}

\author{Xi Chen}
\affiliation{Massachusetts Institute of Technology}

\author{James M.~LeBeau}
\email{lebeau@mit.edu}
\affiliation{Massachusetts Institute of Technology}

\date{\today}


\begin{abstract}

While multislice electron ptychography can provide thermal-vibration limited resolution and 3D information, it relies on the proper selection of many intertwined experimental and computational parameters.
Here, we outline a theoretical basis for selecting experimental parameters to enable robust ptychographic reconstructions.
We develop a series of physically informed metrics to describe the selection of experimental parameters in multislice ptychography.
Image simulations are used to comprehensively evaluate the validity of these metrics over a broad range of experimental conditions.
We develop two metrics, areal oversampling and Ronchigram magnification, which predict reconstruction success with high accuracy.
Lastly, we validate these conclusions with experimental ptychographic data, and demonstrate close agreement between trends in simulated and experimental data.
Using these metrics, we achieve experimental multislice reconstructions at a scan step of $1.0 \unit{\AA/px}$, enabling large field-of-view ($>\!18\unit{nm}$), data-efficient reconstructions.
These experimental design principles enable the routine and reliable use of multislice ptychography for materials characterization.

\end{abstract}

\maketitle

\section{Introduction}
%

Electron ptychography promises high-resolution phase contrast imaging, applicable to a wide variety of materials science problems.
Ptychography was first described using non-iterative reconstruction algorithms \citep{rodenburg_phase_1989,rodenburg_theory_1992,rodenburg_experimental_1993}.
However, the development of iterative algorithms \citep{rodenburg_phase_2004,rodenburg_hard-x-ray_2007} for ptychographic reconstruction has greatly increased the capability and flexibility of ptychography to handle experimental noise, such as unknown illumination conditions \citep{thibault_probe_2009}, scan noise \citep{zhang_translation_2013}, and incoherence \citep{thibault_reconstructing_2013}.
These improvements make electron ptychography practical for the imaging of 2D materials with sub-angstrom resolution \citep{jiang_electron_2018,chen_mixed-state_2020}.
Additionally, iterative ptychography can be extended to reconstruct 3D objects, using a multislice forward model \citep{tsai_x-ray_2016}. Applied to electrons, this allows the imaging of thick samples, and the resolution of 3D structural features \cite{gao_electron_2017,chen_electron_2021,chen_three-dimensional_2021,sha_deep_2022,gilgenbach_three-dimensional_2023}.

While undoubtedly more flexible than non-iterative algorithms, iterative algorithms are inherently difficult to analyze and optimize.
A successful ptychographic reconstruction requires the proper selection of numerous acquisition parameters (e.g, scan step size, probe defocus, diffraction pixel size) and iterative reconstruction parameters.
Past works have considered parameter selection for single-slice ptychography.
Cao et~al. applied a Bayesian algorithm to optimize reconstruction and experimental parameters \citep{cao_automatic_2022}.
However, while Bayesian optimization is an effective method for parameter selection, it does not elucidate the physical mechanisms which lead to these optimal parameters, limiting the intuition that can be built from these findings.
Additionally, Bayesian optimization is much less suited to the problem of experimental parameter selection, as several human-in-the-loop iterations are required to reach an optimum configuration.
Another approach was taken by Edo et~al., who developed a theoretically-backed metric coupling scan step size and diffraction pixel size in single-slice ptychography \citep{edo_sampling_2013}.
However, this work does not guide the selection of other important acquisition parameters in electron ptychography, such as defocus and probe semi-convergence angle.
These parameters have been found empirically to have a large impact on reconstruction quality \cite{chen_electron_2021,gilgenbach_three-dimensional_2023}.
Optimal parameter selection is especially important for the success of multislice ptychography.
Additionally, there are significant differences between the optimal experimental parameters for single-slice and multislice ptychography.
For instance, multi-slice ptychography reconstructions typically require a much smaller scan step size than is required for single-slice ptychography, and much greater than predicted from theory. 
This and other discrepancies reinforce the need for a study of multislice ptychography in particular.

Here, the effect of experimental parameters on multislice ptychography reconstructions is examined using a combination of experimental data and simulated data with realistic artificial noise.
These results are compared to the predictions of several non-dimensional sampling metrics which characterize a multislice ptychography experiment.
These metrics simplify the design of multislice ptychography experiments, and elucidate the constraints and trade-offs between experimental parameters.
Finally, robust and reliable experimental reconstructions are presented under a large range of experimental conditions, in accordance with these sampling metrics.






\begin{figure*}
    \centering
    \includegraphics[width=6in]{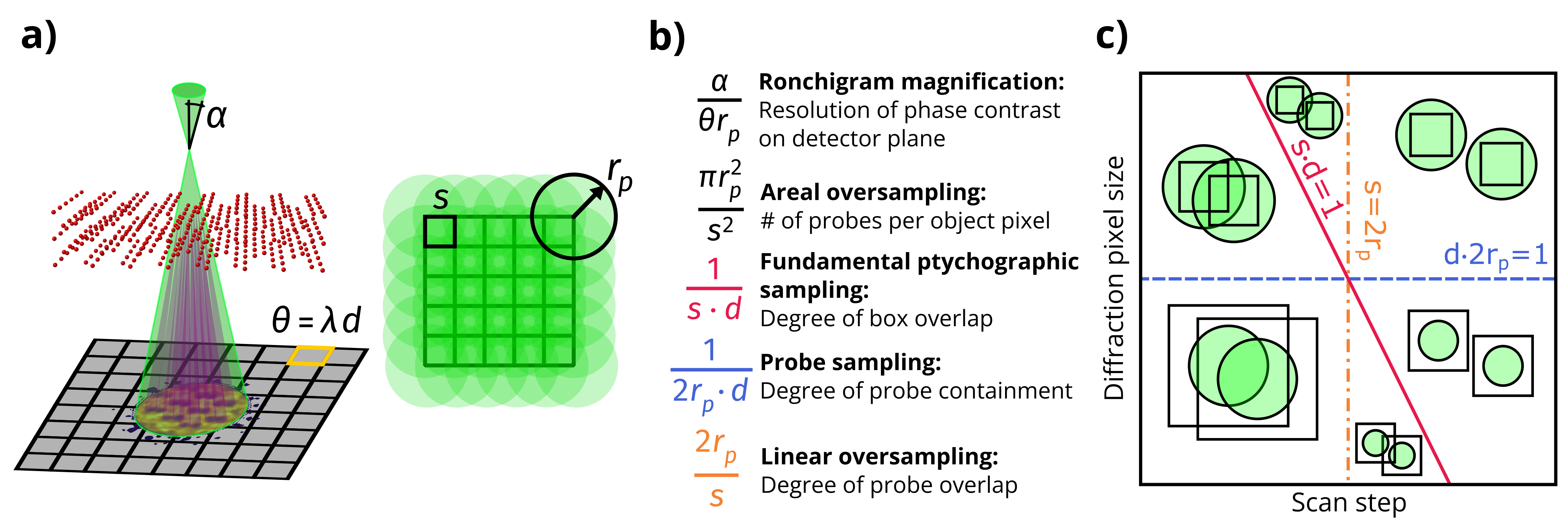}
    \caption{
    a) Schematic of the experimental parameters in multislice ptychography. 
    b) Summary of sampling metrics for ptychography. 
    c) Schematic of the relationships between sampling parameter constraints (F.P.S~- red solid line, linear oversampling - orange dot-dashed line, and probe sampling - blue dashed line). The boxes and green circles represent the sampling box and and probe size.
    See Section \ref{sec:samplingTheory} for definition of the variables.
    }
    \label{fig:1}
\end{figure*}

\section{Theory of sampling metrics}
\label{sec:samplingTheory}

As illustrated in Fig.~\ref{fig:1}a, Ptychography uses a defocused probe to project a convergent beam electron diffraction (CBED) pattern with convergence angle, $\alpha$, onto a pixelated detector.  The selected defocus and convergence angle lead to a representative probe radius $r_p$ [$\unit{\AA}$].
Typically, this probe is spatially scanned, for example, on a square grid with sampling $s$ [$\unit{\AA/px}$],  to remove phase ambiguity.  At each scan position, a diffraction pattern is collected with reciprocal space sampling $\theta$ [$\unit{mrad/px}$] or $d$ [$\unit{\AA^{-1}/px}$].



Successful ptychographic experiments relies on proper sampling of both real and reciprocal space.
Edo et~al. develop a sampling criterion which they term the ``Fundamental ptychographic sampling'' (F.P.S.), which attempts to delineate the sampling in real and reciprocal space necessary to attain a unique object \citep{edo_sampling_2013}.
Each measurement on the detector corresponds to a sampling of reciprocal space with sampling $d$ and extent $N d$, where $N$ is the number of detector pixels in each dimension.
Through a discrete Fourier transform, this corresponds to a real space sampling of $1/N d$ with extent $1/d$.
From the Nyquist sampling theorem, the extent $1/d$ is the size in real space which may be unambiguously reconstructed from a discrete wavefield (amplitude and phase) sampled on the detector plane.
The F.P.S. is the ratio between this real-space sampling window $1/d$ and the real-space scan step size $s$:
\begin{equation}
    \mathrm{F.P.S.} = \frac{1}{s \cdot d}
\end{equation}

An F.P.S. of 1 corresponds to the condition where the sampling box corresponding to each probe position touches but does not overlap.
It is expected that an F.P.S. of at least 2 is required to allow both phase and amplitude reconstruction, allowing an unambiguous solution of the phase problem in general. \citep{edo_sampling_2013}
This criterion was inspired by a widely-used criterion in coherent diffractive imaging, wherein reciprocal space was required to be sampled at two times the object's extent .\citep{rodenburg_phase_1989,miao_phase_1998}
The F.P.S. relaxes this criterion for ptychography, so that the sampled region from a probe position must only overlap (in real-space) with the regions sampled by other probe positions, rather than cover the object's complete extent as in coherent diffractive imaging.

Another metric for successful reconstruction measures the degree of probe overlap in real space, which expresses the constraint that the entire object must be illuminated by at least one probe position.
Electron ptychography experiments typically use a focused or slightly-defocused converged probe in an aberration corrected instrument.
In this geometry, the probe shape is an Airy disk (potentially defocused), and a representative probe radius $r_p$ may be defined as the radius that contains a certain fraction of the incident probe intensity.
This probe radius is then a function of the beam energy, convergence angle, probe defocus, and any optical aberrations present.

Given $r_p$, a dimensionless overlap/oversampling criteria may be defined in one and two dimensions:
\begin{align}
    \mathrm{Linear\:oversampling:}\; &L = \frac{2 r_p}{s} \\
    \mathrm{Areal\:oversampling:}\; &A = \frac{\pi r_p^2}{s^2}
\end{align}
where $s$ is the scan step size.
The two criteria are directly related as $A = \frac{\pi}{4} L^2$.
An oversampling of $L = 1$ corresponds to the condition where probes are just touching ($L = \sqrt{2}$, $A = \pi/2$ is required for overlap along the diagonals and therefore complete coverage of the object).
The linear oversampling $L$ has been most commonly used in literature, often referred to as the `overlap ratio' \citep{edo_sampling_2013}.
In theory, this small amount of overlap would be sufficient to resolve phase ambiguities and fully reconstruct the object wavefunction \citep{rodenburg_phase_1989, faulkner_movable_2004, rodenburg_ptychography_2008}.
However, much greater overlaps are typically used for practical multislice reconstructions \citep{chen_electron_2021, sha_deep_2022}.
In this large-overlap limit, the areal oversampling, $A$, has the desirable property that it represents the average number of probe positions which `sample' a given position on the object.
In contrast, the linear oversampling has no such obvious interpretation in the context of a 2D scan grid.
Rodenburg 2008 implicitly refers to both parameters, referencing $s = r_p$ ($L = 2$) as speeding ptychographic convergence, and later referencing ``each area of the object is generally illuminated about twice'' ($A \approx 2$) \citep{rodenburg_ptychography_2008}.
For small overlaps, these two parameters are close in magnitude.
However, because this current work focuses on the large overlap case of multislice ptychography, where areal oversampling is more meaningful than linear oversampling, $A$ is preferred rather than $L$ as a sampling metric.

In addition to these existing parameters, two additional metrics are introduced.
First, we consider a metric termed the `probe sampling', $P$.
In contrast to the F.P.S, which defines the minimum sampling necessary to unambiguously encode information from the entire object, the probe sampling defines the minimum sampling necessary to unambiguously encode information from the incident probe.
Probe sampling can be derived as the ratio of the real-spacing sampling window $1/d$ and the probe diameter $2 r_p$:
\begin{equation}
    \mathrm{Probe\:sampling:}\; P = \frac{1}{2 r_p d}
\end{equation}
This parameter describes the degree to which the incident probe is contained in the real-space sampling window. It is analogous to the F.P.S., but with the real-space sampling window $1/d$ compared to the probe diameter rather than the scan step size.
In multislice ptychography, containment is desirable to prevent probe self-interference during inter-slice Fresnel propagation in reconstruction.
This requirement may potentially be relaxed by upsampling the measured diffraction patterns prior to reconstruction, increasing the real-space sampling window at the expense of greatly increased computation time.

A final metric considers the phase-contrast information present on the detector.
This metric is termed the `Ronchigram magnification', $M$.
When a sample is imaged with a coherent electron probe of relatively-large convergence angle, overlap and interference between direct and diffracted waves encodes phase contrast information in the direct beam disc.
This interference pattern is termed the Ronchigram \citep{lin_reconstruction_1986,lupini_electron_2011} or Gabor in-line hologram \citep{cowley_reconstruction_1981,lin_reconstruction_1986,cowley_twenty_1992}.
When defocused, the Ronchigram forms a `shadow image' of the sample which contains a mix of real- and reciprocal- space information \citep{cowley_coherent_1979}.
The Ronchigram magnification estimates the size of this shadow image on the detector plane, in terms of detector pixels per distance in real space:
\begin{equation}
    \mathrm{Ronchigram\:mag:}\; M = \frac{\alpha}{\theta \cdot r_p} = \frac{\alpha}{\lambda d \cdot r_p} \propto P \alpha
\end{equation}
where $\alpha$ is the probe semi-convergence angle, and $\theta$ is the reciprocal space sampling in angular units.
For a shadow image, reciprocity equates a point spread function on the detector to a source with finite size.
Ronchigram magnification is therefore the conversion between this point spread function in detector pixels to an effective incoherence in real-space.
A pattern with high Ronchigram magnification has phase information well separated on the detector plane, and thus preserved in the dataset.
On the other hand, a pattern recorded with low Ronchigram magnification fails to capture high spatial frequencies in real space, decreasing contrast and phase information.
$M$ has dimensions of inverse length, but may be non-dimensionalized using the wavelength $\lambda$.
In the interest of physical interpretability, however, $M$ is presented in units of $\unit{px/\AA}$.
It should also be noted that the Ronchigram magnification $M$ is directly proportional to the probe sampling $P$ times probe semi-convergence angle $\alpha$.

\subsection{Summary of metrics}

For convenience and clarity, the sampling metrics discussed above are concisely presented in Fig.~\ref{fig:1}b. The relationship between the different metrics are preseted in Fig.~\ref{fig:1}c, where the central point corresponds to $1/d = s = 2 r_p$, and the sampling window, probe diameter, and scan step are all the same size. The six surrounding regions represent different overlap conditions.
Conditions to the left of the solid red line obey the fundamental ptychographic sampling constraint, meaning the sampling boxes overlap.
Conditions to the left of the dot-dashed orange line obey the linear oversampling constraint, meaning probe positions overlap.
Conditions below the dashed blue line obey the probe sampling constraint, meaning the probe is entirely contained in the sampling box.

The linear overlap and fundamental ptychographic sampling constraints represent hard limits; successful reconstructions are not possible without satisfying these constraints.
The probe sampling constraint is also essential in the absence of detector upsampling, limiting successful reconstructions to the lower left quadrant.
This renders the fundamental ptychographic sampling necessary but not sufficient for proper sampling.
In contrast, Ronchigram magnification does not impose a hard cut off, but represents a gradual loss of phase information in the diffraction plane.

\section{Materials \& Methods}

Simulations of 4D-STEM datasets were performed using the multislice algorithm \citep{cowley_scattering_1957} as described in Kirkland 2010 \citep{kirkland_advanced_2010}.
Simulations of Si \hkl[110] were performed with a $9.2 \times 9.2 \unit{nm}$ supercell that was $20.0 \unit{nm}$ thick.
Simulations were performed with 80 thermal configurations, using a mean-squared displacement $\left<u^2\right> = 5.941\times 10^{-3} \unit{\AA^2}$ \citep{flensburg_lattice_1999}.
To account for the probe effective source size, probe positions were randomly displaced for each thermal configuration based on a $0.5 \unit{\AA}$ full-width half-maximum (FWHM) Gaussian distribution.
Scan position error was modeled using $0.2 \unit{\AA}$ RMS Brownian ($1/f^2$) noise.

Prior to each ptychographic reconstruction, Poisson noise was first added to raw 4D-STEM dataset, and then convolved with a $0.75 \unit{px}$ Gaussian point-spread function (PSF).
The Poisson noise accounts for electron shot noise, and unless otherwise specified corresponds to a dose of $3\times 10^5 \unit{e^-/\AA^2}$.
The Gaussian PSF was chosen to approximate that of an EMPAD detector at $200 \unit{keV}$ \citep{tate_high_2016}.
Each simulated dataset was reconstructed three ways: Using every scan position, using every second scan position, and using every third scan position. This enabled more efficient variation of scan step size.
Data was simulated for a range of parameters: scan step size $0.3$--$1.0 \unit{\AA/px}$, diffraction pixel size $0.6$--$2.0 \unit{mrad/px}$, convergence angle $15$--$30 \unit{mrad}$, and probe defocus $0$--$50 \unit{nm}$ (all in the overfocused condition). Simulations were performed at a beam energy of $200 \unit{keV}$.

For experiments, Si \hkl[110] samples were prepared using cross-section wedge polishing and $\mathrm{Ar}^+$ ion milling, with a final step at  $0.1 \unit{keV}$.
Samples were kept under vacuum to minimize the formation of an oxide layer.
A Thermo Fisher Scientific Themis Z S/TEM was used to collect the experimental data.
4D STEM datasets were collected using a $128\times128$ pixel EMPAD detector \citep{tate_high_2016} with doses ranging from $5 \times 10^5 \unit{e-/\AA^2}$ to $18 \times 10^5 \unit{e-/\AA^2}$.
To minimize the effect of sample damage and contamination, no more than three datasets were taken in a given region of the sample.
The sample was approximately $20 \unit{nm}$ thick in the regions analyzed, as measured using position-averaged convergent beam electron diffraction (PACBED) patterns \citep{lebeau_position_2010}.

As with the simulated data, each dataset was reconstructed using three multiples of scan step size. This leads to further variations in the `effective' doses for reconstruction.
Experimental datasets were collected with a range of parameters: scan step size $0.17$--$1.04 \unit{\AA/px}$, diffraction pixel size $0.60$--$1.20 \unit{mrad/px}$, and probe defocus $0$--$40 \unit{nm}$ (overfocused). Experiments were performed at a beam energy of $200 \unit{keV}$ and convergence angle of $20.2 \unit{mrad}$.
A table of experimental datasets with scan step size, diffraction pixel size, beam current, and dose has been provided in the supplemental material.

Ptychographic reconstructions were performed using the \code{fold\_slice} \citep{chen_three-dimensional_2021} fork of the PtychoShelves software package \citep{wakonig_ptychoshelves_2020}.
Odstrčil et al.'s GPU-accelerated maximum-likelihood/least-squares solver was used as the multislice reconstruction engine \citep{thibault_maximum-likelihood_2012, thibault_reconstructing_2013, tsai_x-ray_2016}.
Datasets from simulation and experiment were both reconstructed using 25x $1 \unit{nm}$ thick object slices.
$8$ incoherent probe modes were used.
Diffraction patterns were padded to $256\times256$ pixels prior to the final reconstruction engine.
A complete listing of reconstruction parameters for data from simulation and experiment are provided in the supplemental material.

Several of the metrics (oversampling, probe sampling, and Ronchigram magnification) require a representative incident probe radius $r_p$.
This was calculated as the radius containing $50\%$ of the incident probe intensity. 
For large defocus values, this is approximately twice the radius predicted from geometric optics $r_{geom} = \alpha \cdot \mathrm{df}$, where $\alpha$ is the probe semi-convergence angle in radians and $\mathrm{df}$ is the probe defocus.
Probe radii as a function of beam energy, convergence angle, and probe defocus are provided in the supplementary material.

To evaluate the above metrics, multislice simulations and multislice ptychographic reconstructions were performed with varying convergence angles, scan step sizes, diffraction pixel sizes, and defocus values.
In total, this encompassed 632 reconstructions.
For each reconstruction, quality was then assessed by comparing the projected phase of the reconstruction to a thermally-averaged, projected transmission function used for the multislice method.
This ground-truth phase was additionally bandwidth-limited to a maximum resolution of $4.0 \unit{\AA^{-1}}$, corresponding to the maximum theoretical resolution from a detector with diffraction pixel size $1.6 \unit{mrad/px} = 0.064 \unit{\AA^{-1}/px}$.
The mean-squared error (MSE) between the reconstructed and ground-truth phase was used to indicate the quality of reconstruction.
In the following, this error metric is referred to as the ``real-space MSE'', in contrast to the ``Fourier MSE'' which is optimized by the multislice ptychography algorithm.
This implies real-space MSE is sensitive to overfitting in the Fourier domain caused by poor regularization or a poor noise model.

\section{Results \& Discussion}

\begin{figure}
    \centering
    \includegraphics{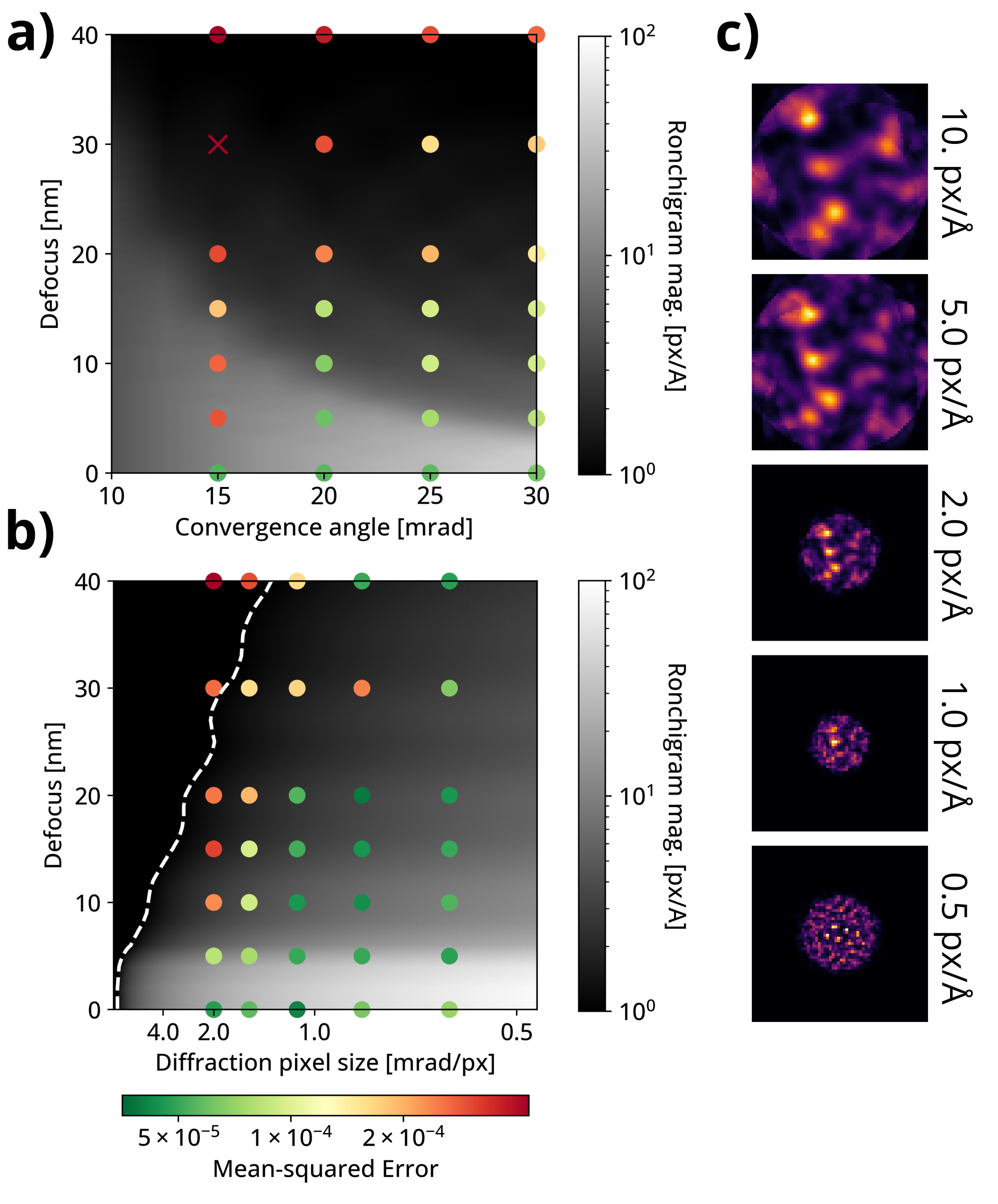}
    \caption{
    a) Map of Ronchigram magnification and simulation results as a function of probe defocus and diffraction pixel size.
    Data was taken at a scan step size of $0.3 \unit{\AA/px}$ and convergence angle of $25 \unit{mrad}$.
    Dotted white line indicates a Ronchigram magnification of $1 \unit{px/\AA}$. Crosses indicate reconstructions which failed to converge.
    b) Map of Ronchigram magnification and simulation results as a function of probe defocus and convergence angle.
    Data was taken at a scan step size of $0.3 \unit{\AA/px}$ and diffraction pixel size of $1.5 \unit{mrad/px}$ = $0.060 \unit{\AA^{-1}/px}$. Crosses indicate reconstructions which failed to converge.
    c) Representative CBED patterns at Ronchigram magnifications between $0.5 \unit{px/\AA}$ and $10. \unit{px/\AA}$. Patterns are shown for a $20 \unit{nm}$ thick sample, and are displayed with a linear scale.
    }
    \label{fig:2}
\end{figure}

The simulation results indicate that Ronchigram magnification correlates well with reconstruction quality, as shown in Fig.~\ref{fig:2}a for varying diffraction pixel size and defocus (constant convergence angle).
Small defoci are close to an `infinite magnification' condition, leading to high Ronchigram magnifications and well-separated phase information on the detector plane.
At higher defoci, more phase variation is present on the detector plane.
With coarser diffraction sampling, this phase variation is sampled less well.
In either scenario, phase information is gradually lost.
Based on the reconstruction quality in Fig.~\ref{fig:2}a, an approximate cutoff magnification of $1 \unit{px/\AA}$ (dashed white contour) is required for a high-quality reconstruction.
Fig.~\ref{fig:2}b shows the same dataset but with a varying convergence angle rather than diffraction pixel size.
A general trend of increasing reconstruction quality with increased convergence angle is observed.
This is consistent with the prediction of Ronchigram magnification.
Moreover, for different amounts of Ronchigram magnification, Fig.~\ref{fig:2}c, shows the contrast achieved on the detector plane.
At low magnifications, fine detail is lost in the central patterns. This detail is better preserved at higher magnifications.

The influence of probe sampling on the same  reconstruction data from simulation is provided in the supplementary material.
At a single convergence angle, the metrics of probe sampling and Ronchigram magnification are proportional and therefore indistinguishable.
When convergence angle is varied, however, the probe sampling required for a successful reconstruction changes while the required Ronchigram magnification remains approximately constant, as shown in the supplementary material.
Therefore, Ronchigram magnification better encapsulates the effect of convergence angle on  reconstruction fidelity.
Nevertheless, it is possible that there are additional factors which cause larger convergence angles to reconstruct better besides increased Ronchigram magnification.
For instance, larger convergence angles are able to `fill out' reciprocal space better in the presence of limited dose.

\begin{figure}
    \centering
    \includegraphics{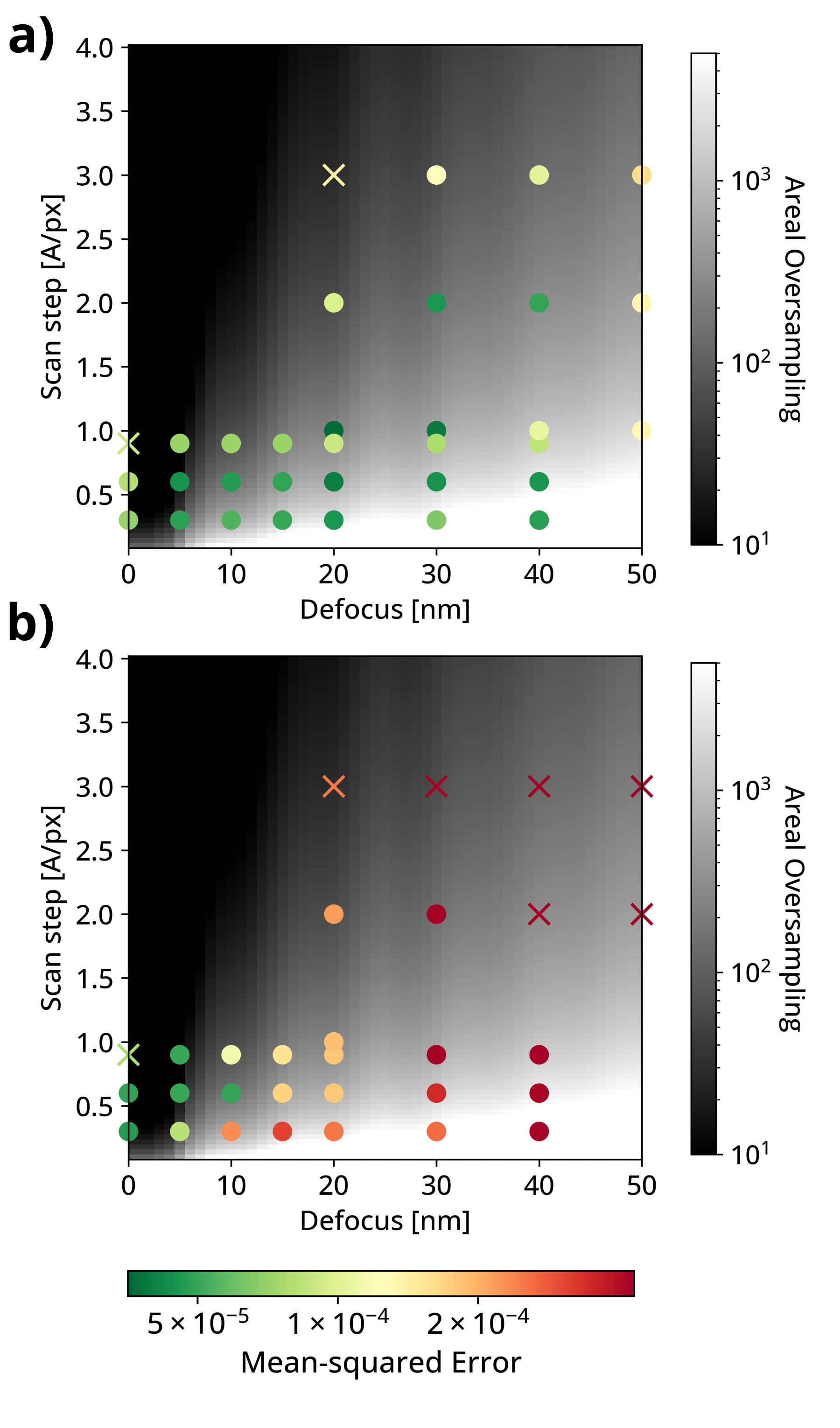}
    \caption{
    Maps of areal oversampling and simulation results as a function of probe defocus and scan step size, at a constant beam energy of $200 \unit{keV}$, convergence angle of $25 \unit{mrad}$. 
    a) corresponds to a diffraction pixel size of $0.6 \unit{mrad/px}$, while b) corresponds to a diffraction pixel size of $2.0 \unit{mrad/px}$.
    Crosses indicate reconstructions that failed to converge.
    }
    \label{fig:3}
\end{figure}

While Ronchigram magnification describes coherence effects on the detector plane, areal oversampling describes sampling in real space. The influence of areal oversampling on reconstruction quality is shown as a function of defocus and scan step size in 
Fig~\ref{fig:3}.
At a diffraction pixel size of $0.6 \unit{mrad/px}$ (Fig.~\ref{fig:3}a), areal oversampling is a good proxy for reconstruction quality. Acceptable reconstructions are obtained even at the highest scan step size tested, $3.0 \unit{\AA/px}$.
At a higher diffraction pixel size ($2.0 \unit{mrad/px}$, Fig.~\ref{fig:3}b), the detector plane effects dominate and penalize higher defocus values.
In these conditions, reconstruction error is minimized by small defocus.
Consistent with past investigations of multislice ptychography, high oversampling is required for the best quality reconstructions \citep{chen_electron_2021,sha_deep_2022}.
This is especially true at larger diffraction pixel sizes and can be rationalized through either the F.P.S. or Ronchigram magnification metrics.

\begin{figure}
    \centering
    \includegraphics[width=3.2in]{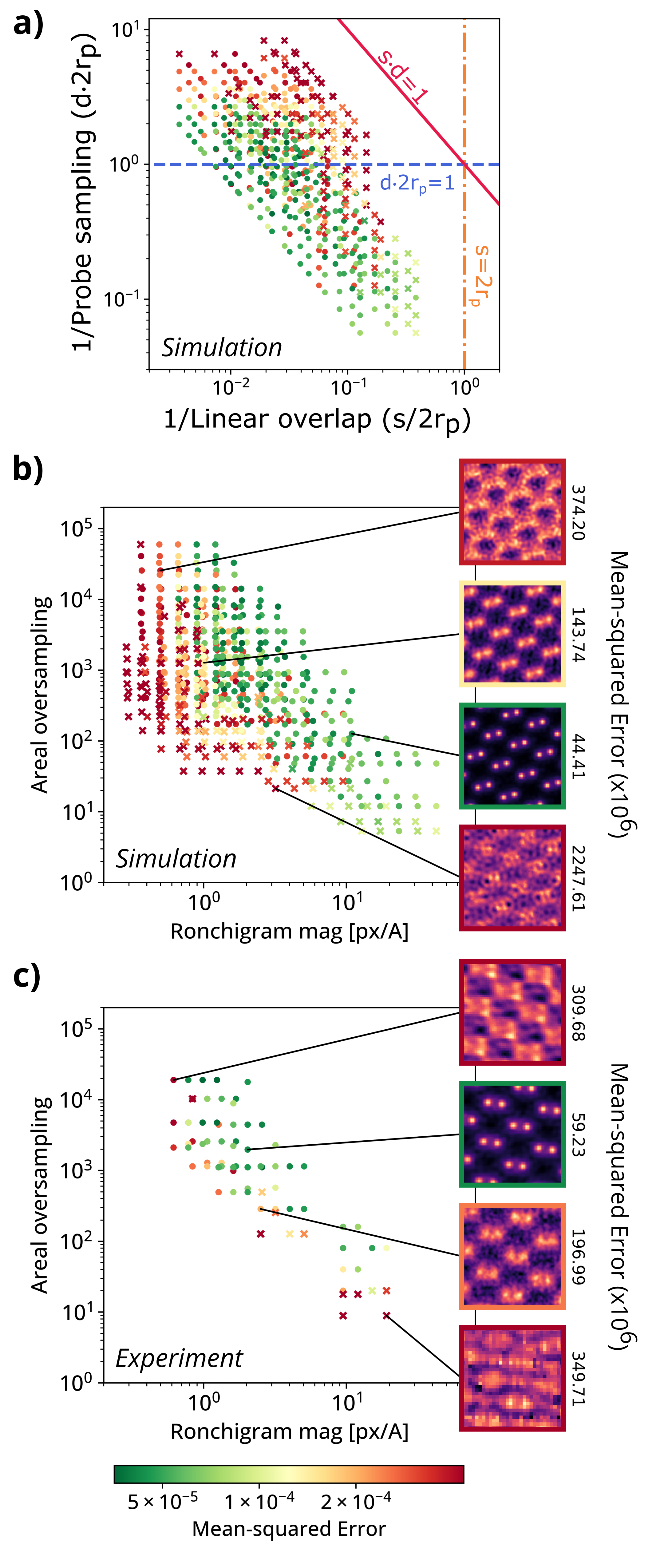}
    \caption{
    a) Simulated data reconstruction quality   projected onto the metrics of probe sampling and linear overlap. Overlaid are the theoretical constraints of probe sampling (dashed blue), linear overlap (dash-dotted orange), and fundamental ptychographic sampling (solid red).
    b) Simulated data reconstruction quality   projected onto the metrics of areal oversampling and Ronchigram magnification. Reconstructed object phase and mean-squared error are shown for selected data points.
    c) Experiment projected in the same manner as b).
    }
    \label{fig:4}
\end{figure}

To evaluate the power of metrics to describe reconstruction success in a wide range of experimental conditions, all datapoints are projected onto subspaces of the metrics.
The reconstructions are shown projected onto the linear overlap and probe sampling metrics in Fig.~\ref{fig:4}a, along with the three geometric sampling constraints described in Fig.~\ref{fig:1}c.
Broadly, lower probe sampling is correlated with poor reconstructions.
This trend, however,  appears to vary significantly with convergence angle, and failures occur with sampling significantly higher than predicted by either the linear overlap or fundamental ptychographic sampling metric.

The same dataset is shown projected onto the areal oversampling and Ronchigram magnification metrics (Fig.~\ref{fig:4}b).
At representative data points, the reconstructed phase and mean-squared error are shown to the right.
Reconstructions with Ronchigram magnification under approximately $1 \unit{px/\AA}$ are generally unsuccessful.
Unlike with the probe sampling metric, this cutoff value does not appear to change as a function of convergence angle.
As a result, at moderate Ronchigram magnifications, large areal oversampling is required for successful reconstruction.
At higher magnifications, however, successful reconstructions can be achieved even with relatively low areal oversampling.
Empirically, Ronchigram magnification and areal oversampling are the two parameters which resulted in the most complete separation of successful and unsuccessful reconstructions. Other parameters, however, may be dominant in different regions of parameter space.


To validate the conclusions drawn from the simulated dataset, the data collected in experiments using similar conditions are reconstructed and presented in Figure~\ref{fig:4}c.
Trends similar to the simulation results are observed, indicating that simulations are effectively capturing the limiting factors in  data from experiments.
It should be noted that these experiments have differing effective sample doses, as detailed in the methods section.


\begin{figure*}
    \centering
    \includegraphics{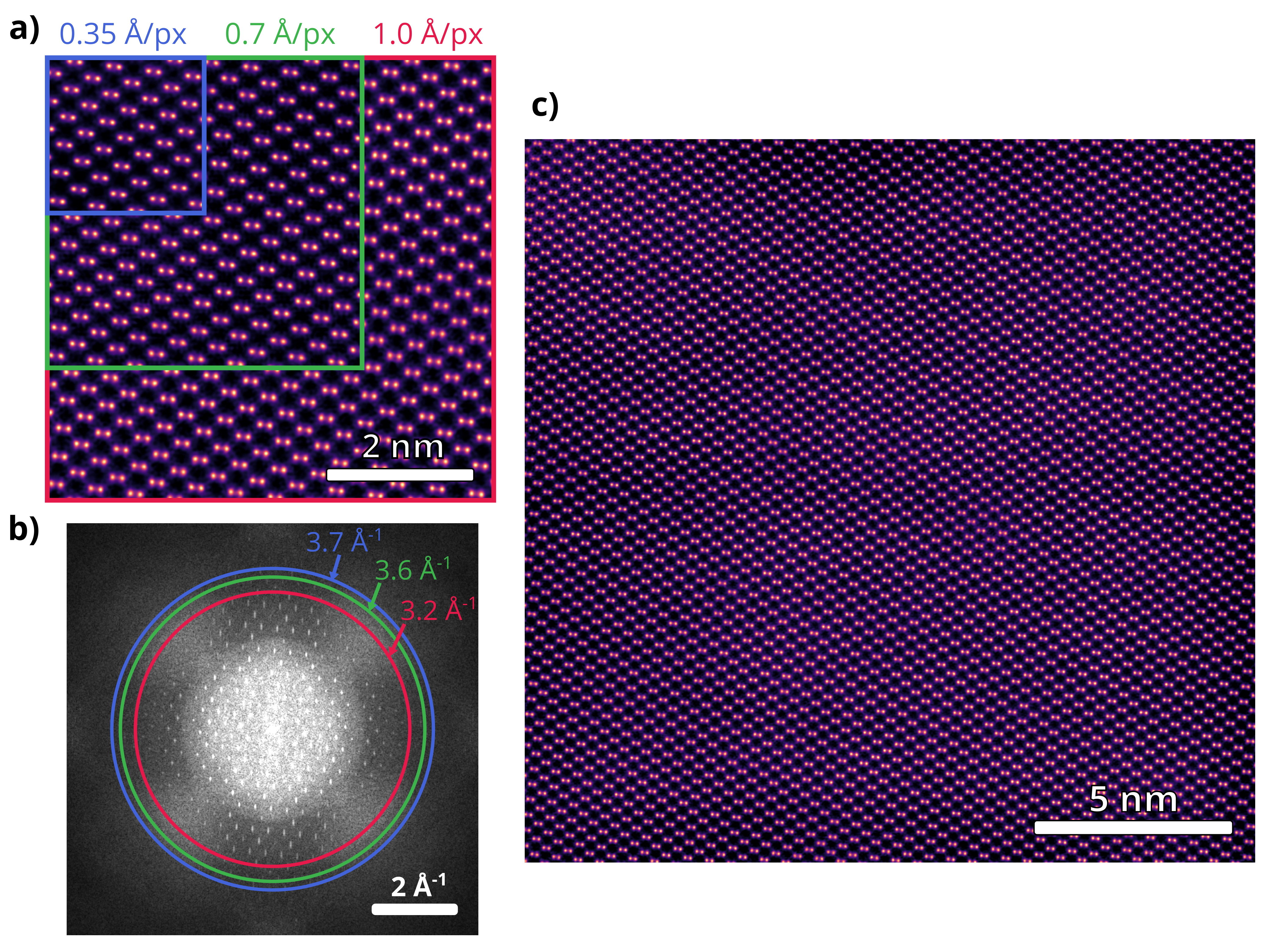}
    \caption{
    a) Reconstructions from $60 \times 60$ probe positions taken at three scan step sizes.
    b) Discrete Fourier transform (DFT) of the $0.35 \unit{\AA/px}$ reconstruction, overlaid with the achieved resolutions of each dataset. A Hann window was applied to the object prior to the DFT, and the pattern was displayed with a gamma of $0.2$.
    c) Reconstruction from $180 \times 180$ probe positions at $1.0 \unit{\AA/px}$.
    }
    \label{fig:5}
\end{figure*}

Applying these metrics, high-quality reconstructions are routinely possible in simulation \textit{and} experiment.
For example, reconstructions of the datasets from the \hkl[110] Si experiments, Figure~\ref{fig:5}a,b, are shown for three different  scan step sizes, but a constant number of scan pixels. Thus, the data size and and acquisition time are constant, but have different fields of view. Moreover, the dose is also constant at $50{\small,}000 \unit{e/\AA^2}$, as measured on the detector.

While multislice reconstructions in literature tend to use very small scan step sizes, around $0.3-0.4 \unit{\AA/px}$ \cite{chen_electron_2021,sha_deep_2022,gilgenbach_three-dimensional_2023}, larger scan steps are possible with minimal loss in reconstruction quality as shown by the resolutions in Fig.~\ref{fig:5}b. 
This allows for much larger fields of view to be reconstructed and for higher data efficiency.
Figure~\ref{fig:5}c, for example, is a reconstruction with $1.0 \unit{\AA/px}$ scan step size and 180x180 scan pixels.
From 2.1 GiB of raw detector data, 191 MiB of object data is reconstructed. This corresponds to approximately 11 bits of raw data for every bit of reconstructed data.
In comparison, a dataset with the same camera length but a scan step size of $0.35 \unit{\AA/px}$ results in 37.5 MiB of object data, for a data redundancy of 57x.
Tuning these parameters therefore provides a method to `use every byte' of data in experiment.
For detectors which are limited primarily by data transfer speed, this allows data to be collected faster and enables the routine use of multislice ptychography for a broader range of applications.

\section{Conclusions}

This work demonstrates that reconstruction quality in multislice electron ptychography can be predicted by two metrics: Areal oversampling and Ronchigram magnification.
A theoretical basis is given for each of these metrics, providing intuition into success and failure in multislice ptychography.
More than exact experimental parameters, this provides a terrain for optimizing ptychographic experiments on a specific material and for a specific purpose.
Successful reconstructions are demonstrated over a large range of experimental conditions, and a close match is found between the reconstruction convergence behavior of simulation and experiment.
With optimized metrics, large field-of-view reconstructions are demonstrated in experiment with high data-efficiency.
This work demonstrates the robust and reliable use of multislice ptychography, enabling routine application to a broad class of applications and analyses in materials science.

\section{Acknowledgments}

The authors acknowledge support from the Air Force Office of Scientific Research (FA9550-20-0066) and the Department of Homeland Security (22CWDARI00046-01-00).
This work was performed with the assistance of MIT SuperCloud and was carried out in part through the use of MIT's Characterization.nano facilities.
Data processing was carried out in part using the cSAXS ptychography MATLAB package developed by the Science IT and the coherent X-ray scattering (CXS) groups, Paul Scherrer Institut, Switzerland.
The version of this package used contains modifications and improvements by Zhen Chen, Yi Jiang, as well as the authors.

\section{Data Availability}

A subset of simulated and experimental data is available through Zenodo.  
Datasets are provided in the EMPAD `.raw' file format, which is a raw binary file format consisting of 32-bit floating-point numbers.
Python and Matlab scripts are provided to read and write this file format.
For each dataset, metadata is provided in Javascript object notation (JSON) format.
A schema and description of this metadata format is also provided, with the goal of enabling easier interchange of 4D-STEM datasets and interoperability between reconstruction and analysis tools. A modified version of the \code{fold\_slice} MATLAB package is provided, as well as instructions on running this package using the provided data.

Other data and code is available upon reasonable request.

%

\bibliography{references}

\end{document}